\documentclass[10pt, conference, letterpaper]{IEEEtran}
\IEEEoverridecommandlockouts
\usepackage{cite}
\usepackage{amsmath,amssymb,amsfonts}
\usepackage{algorithmic}
\usepackage{graphicx}
\usepackage{textcomp}
\usepackage{xcolor}
\usepackage{amsthm,amssymb}
\usepackage{epstopdf}
\usepackage{url}
\usepackage{wrapfig}

\ifCLASSOPTIONcompsoc
\usepackage[caption=false,font=normalsize,labelfon
t=sf,textfont=sf]{subfig}
\else
\usepackage[caption=false,font=footnotesize]{subfig}
\fi
\usepackage[algoruled, vlined, linesnumbered]{algorithm2e}  
\usepackage{booktabs}

\newcommand{\ind}{1{\hskip -2.5 pt} \mathrm{I}}

\SetKwInOut{Parameter}{Parameters}

\makeatletter
\def\endthebibliography{%
	\def\@noitemerr{\@latex@warning{Empty `thebibliography' environment}}%
	\endlist
}
\makeatother

\DeclareMathAlphabet{\mathcalboondox}{U}{BOONDOX-calo}{m}{n}
\SetMathAlphabet{\mathcalboondox}{bold}{U}{BOONDOX-calo}{b}{n}
\DeclareMathAlphabet{\mathbcalboondox}{U}{BOONDOX-calo}{b}{n}

\DeclareMathOperator*{\argmin}{arg\,min}

\def\BibTeX{{\rm B\kern-.05em{\sc i\kern-.025em b}\kern-.08em
		T\kern-.1667em\lower.7ex\hbox{E}\kern-.125emX}}

\begin{document}
	
	\title{Energy Savings under Performance Constraints via Carrier Shutdown with Bayesian Learning}
	
	\author{\IEEEauthorblockN{Lorenzo Maggi$^\dagger$, Claudiu Mihailescu$^*$, Qike Cao$^*$, Alan Tetich$^*$, Saad Khan$^*$, Simo Aaltonen$^*$, Ryo Koblitz$^\dagger$,\\Maunu Holma$^*$, Samuele Macchi$^*$, Maria Elena Ruggieri$^*$, Igor Korenev$^*$, Bjarne Klausen$^*$}
		\IEEEauthorblockA{\textit{Nokia (Bell Labs$^\dagger$ \& Mobile Networks$^*$)} 
		}
	}
	
	\maketitle

\begin{abstract}
By shutting down frequency carriers, the power consumed by a base station can be considerably reduced. However, this typically comes with traffic performance degradation, as the congestion on the remaining active carriers is increased.

We leverage a hysteresis carrier shutdown policy that attempts to keep the average traffic load on each sector within a certain min/max threshold pair. We propose a closed-loop Bayesian method optimizing such thresholds on a sector basis and aiming at minimizing the power consumed by the power amplifiers while maintaining the probability that KPI's are acceptable above a certain value. 
We tested our approach in a live customer 4G network. The power consumption at the base station was reduced by $11\%$ and the selected KPI's met the predefined targets.
\end{abstract}


\section{Introduction}

As new mobile network generations are rolled out, the energy required to transmit over the air per unit of information ($J/\mathrm{bit}$) tends to decrease. This is mainly thanks to the increased energy efficiency of the hardware deployed at the base station, as well as to the design of better resource management algorithms. For instance, with respect to its predecessors, 5G better focuses transmitted energy towards users via analog beamforming, allows multiple transmissions to multiple users to occur at the same via massive MIMO (Multiple-Input-Multi-Output) spatial multiplexing, and reduces signaling overhead by lean carrier design \cite{lopez2022survey}. 
However, such advances alone prove to be insufficient to curb energy consumption at the base station and keep up with the confluence of increased traffic volume, skyrocketing energy costs, and more stringent environmental regulations. 
Hence, the telecommunication industry is striving to find new ways to reduce the carbon footprint of its networks by using existing resources parsimoniously. 

It is well known that power amplifiers (PA) are the main source ($>65\%$, \cite{lopez2022survey}) of power consumption at radio frequency (RF) in a base station. Thus, a good practice for reducing consumption at the base station is to activate as few PA's as possible, while not (overly) degrading network performance. 

Different resource management techniques leading to PA switch-off operate on different time scales and domains (frequency and/or antennas). 
One of such techniques is \emph{symbol-level shutdown} (also called \emph{cell-DTX} in LTE \cite{frenger2011reducing}) which deactivates BS hardware components in the absence of traffic and operates on a time scale of tenths of milliseconds. 
Its main advantage is its negligible impact on traffic performance; however, if the number of users is sufficiently high, the chance of observing periods with no traffic is small. 

A second option to turn off hardware circuitry and reduce consumption is deactivating a certain number of antennas. By doing so, the rank of the transmission channel decreases as well as the number of available layers (i.e., the number of streams over which simultaneous communication can occur). This finally leads to a throughput decrease.

\begin{figure}[h]
	\setlength\abovecaptionskip{0.2\baselineskip}
	\centering
	\includegraphics[width=\linewidth]{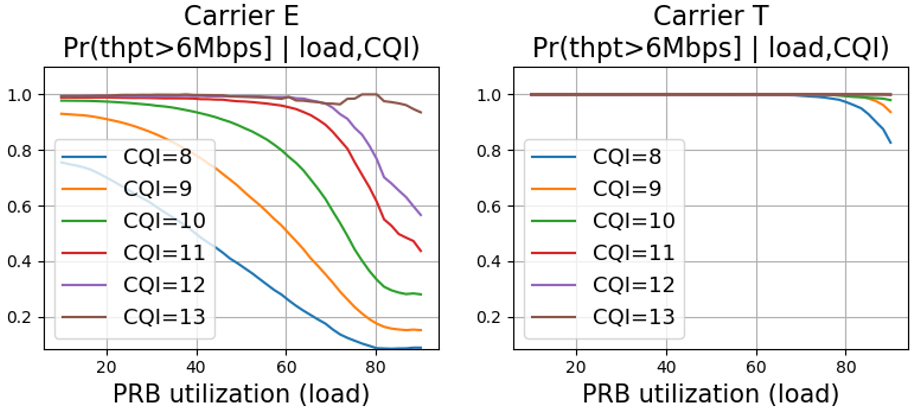}
	\caption{(Live network data) probability that downlink throughput exceeds $6$ Mbps on LTE layer E (freq. 800 MHz, bandwidth 10 MHz) and layer T (freq. 1800 MHz, band 20 MHz) versus PRB utilization $[\%]$ and CQI.}
	\label{fig:thpt_vs_load_CQI}
\end{figure}

In this work we adopt a third option for energy savings, consisting in shutting down frequency carriers. This allows PA's to be switched off over longer time periods, in the order of tens of seconds to few minutes. 
\begin{wrapfigure}{r}{0.2\textwidth}
	\setlength\abovecaptionskip{-0.7\baselineskip}
	\vspace{-1em}
	\begin{center}
		\includegraphics[width=\linewidth]{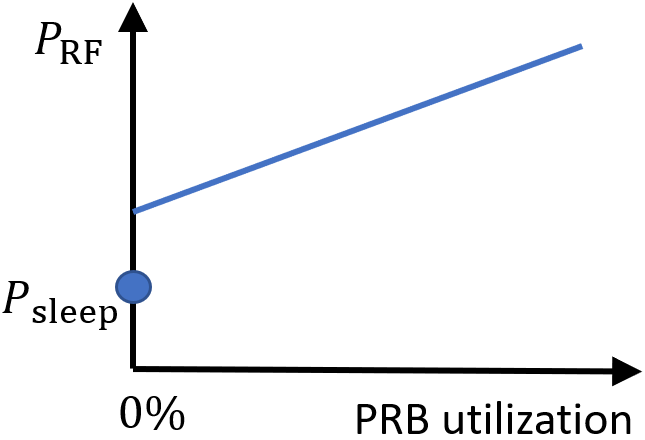}
	\end{center}
	\caption{(Stylized) power consumption of power amplifier versus PRB utilization.}
	\vspace{-.6em}
	\label{fig:P_vs_load}
\end{wrapfigure}
Upon a carrier shutdown, user traffic and signaling transfer to the remaining active carriers(s). Thus, the load on the remaining carriers increases, and the traffic performance typically degrades (see Figure \ref{fig:thpt_vs_load_CQI}). It is known \cite{li2020power} that the power consumed by the PA can be well approximated by an affine function of the kind $P_{\mathrm{RF}}(\ell)=a\ell +b$ of the resource utilization rate $\ell$ for $\ell> 0$. However, $P_{\mathrm{RF}}$ presents a discontinuity at $\ell=0$ ($P_{\mathrm{RF}}(0)=P_{\mathrm{sleep}}<b$, see Figure \ref{fig:P_vs_load}). Hence, the energy increment due to the load increase over active carriers is over-compensated by the PA switch-off, which eventually leads to energy savings. 
We remark that a single PA may be associated to different carriers, possibly across multiple technologies (e.g., 4G and NR). So, deactivating a carrier does not necessarily imply that the serving PA is also turned off.

\begin{figure}[h]
	\setlength\abovecaptionskip{-.2\baselineskip}
	\vspace{-.1em}
	\centering
	\includegraphics[width=.9
	\linewidth]{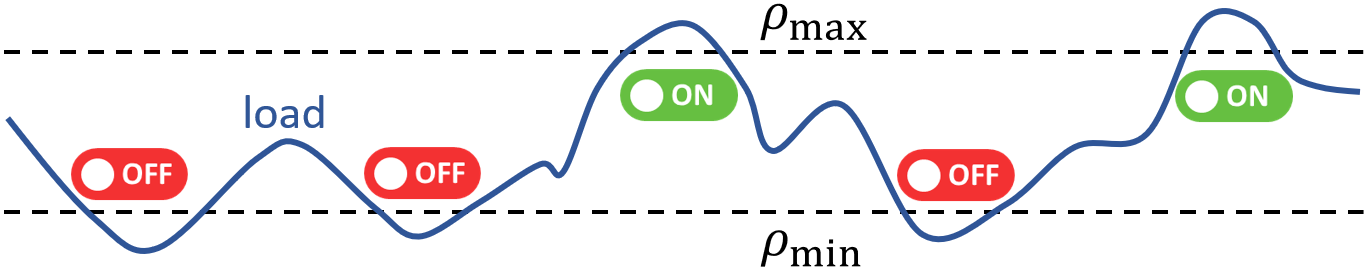}
	\caption{Hysteresis carrier shutdown policy. The average load on the active carriers in the sector is compared against thresholds $\rho_{\min}$ and $\rho_{\max}$ to decide whether to shut down or reactivate a carrier, in a pre-determined order.}
	\vspace{-.5em}
	\label{fig:load_vs_rho}
\end{figure}

\noindent \textbf{Our contribution}. We leverage a method that reduces energy consumption at the base station by shutting carriers down in a pre-defined order (e.g., in decreasing order of frequency). According to a hysteresis mechanism, the next carrier in line is switched off (on, respectively) if the load on the sector is lower (greater, resp.) than a certain min-threshold $\rho_{\min}$ (max-threshold $\rho_{\max}$, resp.). Thus, the load is maintained within the interval $[\rho_{\min};\rho_{\max}]$.
By using an over-the-top architecture, we optimize such thresholds on a sector basis, with the aim of minimizing the energy consumed by the PA's while ensuring that certain KPI's meet pre-defined constraints with high confidence. 
We designed a parametric Bayesian algorithm converging to good threshold values in a handful of iterations and capable of adapting to varying channel conditions.
We validated our method via a live customer 4G network trial during which we reduced the power consumption at the base station by 11$\%$ while meeting the KPI constraints with the pre-defined confidence of $89\%$. \\

\subsection{Related works } 

Carrier shutdown is mentioned as a promising technique for reducing the power consumption at the base station in several recent technological surveys such as \cite{lopez2022survey}, \cite{tan2022energy}, \cite{salahdine2021survey} and industry white papers as \cite{huaweiWP}, \cite{zteWP}. A similar approach allows the base station to adapt the bandwidth to the traffic needs via the concept of \emph{bandwidth part}, without the need of powering off the whole carrier, as described in \cite{kim2020evolution}, \cite{li2020power}.
The work in \cite{alam2015scalable} proposes a method to switch off the entire base station (instead of just carriers) when the load on the base station is sufficiently low. In \cite{feng2017base}, the authors illustrate the challenges of base station deactivation, among which coverage loss is crucial. Finally, the work \cite{tano2019kpi} investigates the impact of different level of hardware sleep state on network performance.
 
The sources above agree on the fact that carrier shutdown should not be performed at the expense of traffic performance over-degradation. 
Yet, to the best of our knowledge, we are the first to design an effective method achieving a satisfying (and configurable by the operator) trade-off between energy consumption and network performance via carrier shutdown. 

\section{Problem formulation} \label{ref:pb_formulation}

Let us consider a base station, where a set of frequency carriers $\mathcal C$ is deployed to serve the mobile users in a specific sector.  
We assume that a subset of the carriers $\mathcal C$ can be shut down at any time, and the corresponding attached users are redirected to the remaining active carriers, whose frequency/time resource utilization consequently increases. This typically leads to a degradation of traffic performance (see Figure \ref{fig:thpt_vs_load_CQI}) as measured by network Key Performance Indicators (KPI's).  
On the other hand, the power consumed by the radio units reduces: the increased consumption in the active carriers due to a higher resource utilization is typically over-compensated by the PA consumption reduction induced by carrier shutdown. 

We now introduce some notation. We call $\mathcal A_t\subset \mathcal C$ the set of active carriers at time $t$. We assume that at least one carrier must be left active at any time, to ensure coverage; hence, $|\mathcal A_t|\ge 1, \ \forall \,t$. 
We denote by $w_t(\mathcal A)$ the power consumed by the PA's serving carriers $\mathcal C$ at time $t$ when carriers $\mathcal A$ are active. We assume that a list of $K$ KPI's is constantly monitored on carriers $\mathcal A'_t\supset \mathcal A_t$ that include the active carriers $\mathcal A_t$ in the sector, and possibly also carriers of neighboring cells that could be negatively impacted by our carrier shutdown policy. 
 
We require that KPI's be jointly acceptable on a each carrier with a desired likelihood $\xi$. To this aim, we define a Boolean function $f(\{\mathrm{KPI}^{i,c}_t\}_{i=1}^K)$ that returns 1 if KPI's are acceptable and 0 otherwise, where $\mathrm{KPI}^{i,c}_t$ is the $i$-th KPI measured at time $t$ on carrier $c\in \mathcal A'_t$. E.g., the most natural way to define $f$ is to set a minimum target level $y$ for each KPI and to require that \emph{each} KPI for a given carrier exceeds its target value, i.e., \vspace{-.8em}
\[
\vspace{-.4em}
f\left(\{\mathrm{KPI}^{i,c}_t\}_{i=1}^K\right) := \bigcap_{i=1}^K \left( \mathrm{KPI}^{i,c}_t\ge y^i \right).
\]
Our goal is to determine, for a specific sector and at any time $t\ge 0$, which carriers $\mathcal A_t$ should be activated to minimize the long-run average power consumed by the PA's whilst ensuring that the selected KPI's are acceptable for at least a portion $\xi$ of the time. More formally, our objective writes: \vspace{-.2em}
\begin{align}
	& \min_{A_t\subset \mathcal C} \lim_{T\rightarrow \infty} \, \frac{1}{T+1} \sum_{t=0}^{T} \mathbb E\left[w_t(\mathcal A_t)\right] \label{eq:es1} \\
	& \mathrm{s.t.} \ \lim_{T\rightarrow \infty} \! \frac{1}{\sum_{t=0}^{T} \!|\mathcal A'_t|} \! \sum_{t=0}^{T} \!\sum_{c\in \mathcal A'_t} \!\mathbb E \left[f\left(\{\mathrm{KPI}^{i,c}_t\}_{i=1}^K\right)\right] \ge \xi. \label{eq:es2}
\end{align} 
where the expectation is with respect to the traffic fluctuations.

Examples of KPI's that one may want to preserve upon carrier shutdown are statistics (e.g., mean or percentile across connected users) of integrity KPI's (e.g., downlink/uplink throughput and traffic volume), mobility KPI's (e.g., inter/intra-frequency handover success rate), accessibility KPI's (e.g., setup-success/drop-call rate), availability KPI's (e.g., cell availability), or a combination of those. 

We finally observe that, as the carrier shutdown activity on a sector may affect the performance on neighboring sectors, one should ideally rewrite \eqref{eq:es1}-\eqref{eq:es2} as a joint optimization problem across different sectors. We justify our choice to decouple the carrier shutdown problem across different sectors by claiming that our impact on inter-cell mobility is limited, since we ensure that at least one carrier (typically, the lowest frequency) is always active in each sector, which ensures good coverage.

\section{Solution architecture} \label{sec:sol_arch}

We here describe the computing architecture of our energy savings via carrier shutdown method. In Section \ref{sec:algo} we will delve into its algorithmic details.

\begin{figure}[h]
	\setlength\abovecaptionskip{0\baselineskip}
	\vspace{-1em}
	\centering
	\includegraphics[width=.85\linewidth]{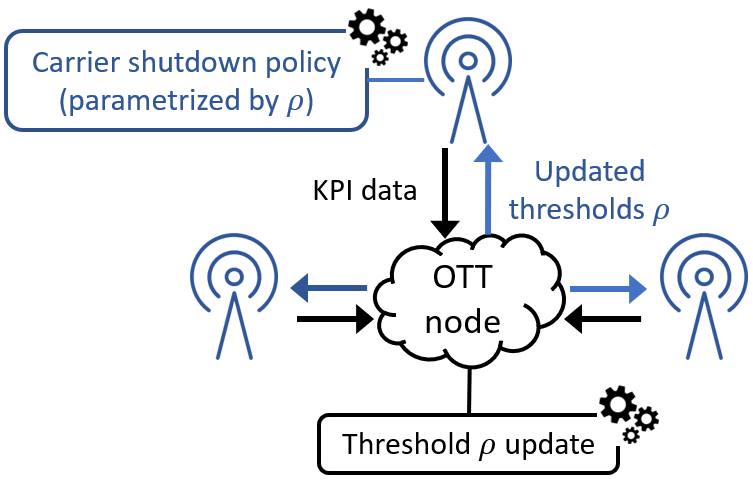}
	\caption{Solution implementation architecture}
	\vspace{-.5em}
	\label{fig:architecture}
\end{figure}

\noindent \textbf{Base station: Carrier shutdown policy implementation.} 
In our solution, the logic handling carrier shutdown is implemented at the base station. We first describe the rationale behind it. Typically, Quality of Service (QoS) is negatively correlated with the Physical Resource Block (PRB) utilization rate (also simply denoted here as \emph{load}) at the base station: the higher the load, the worse the QoS, as shown, e.g., in Figure \ref{fig:thpt_vs_load_CQI}.  
Thus, in order to prevent QoS degradation, one should cap the average load of the active carriers to a certain upper value. On the other hand, energy savings are achieved by shutting carriers down, which eventually leads to a load increase on active carriers; thus, the load should not be kept too low either.  


For such reasons, we use a carrier shutdown policy of hysteresis type, that attempts to keep the average load on active carriers in a sector comprised within $[\rho_{\min};\rho_{\max}]$. When the load is lower than $\rho_{\min}$, then a carrier is shut down; conversely, a carrier is reactivated when the load exceeds $\rho_{\max}$.  

Upon a carrier shutdown decision, the base station gradually reduces its downlink power on the carrier, which forces users to attach to a different carrier or base station.

In our solution, carriers are switched off in a pre-defined order (and back on, in the reverse order) called $\mathcalboondox O$. For instance, a reasonable design choice that preserves network coverage is to shut carriers down in decreasing order of frequency. Indeed, it is known \cite{pi2011introduction} that as the carrier frequency increases, path loss also increases, hence coverage reduces.

The general procedure we used for carrier shutdown is described in Algorithm \ref{alg:LTE1103}, where $\mathcal A_{t}=\{c_{1},\dots,c_{a_{t}}\}$ is the set of active carriers in time period $[t-1,t)$.

\begin{algorithm}
	\KwIn{Sector carriers $\mathcal C=\{c_i\}_{i=1}^{|\mathcal C|}$, sorted in  order $\mathcalboondox O$. Initial set of active carriers $\mathcal A_0$.}
	\Parameter{Load thresholds $\rho_{\min},\rho_{\max}$ ($\rho_{\min}\!\!<\!\!\rho_{\max}$).}
	\For{time instants $t=0,1,\dots$}{
		Compute the average traffic load $\ell_{t}$ on carriers $\mathcal A_{t}$ \\
		\If{$(\ell_{t}<\rho_{\min}) \land (a_{t}>1)$}{Shut down carrier $a_{t}$; set $a_{t+1}=a_{t}-1$.}
		\Else{\If{$(\ell_t>\rho_{\max}) \land (a_{t}<|\mathcal C|$)}{Switch on carrier $a_{t}+1$; set $a_{t+1}:=a_{t}+1$}
		\textbf{else} Set $a_{t+1}:=a_{t}$}
	} 
	\caption{(Vanilla) carrier shutdown policy}
	\label{alg:LTE1103}
\end{algorithm}

We remark that the carrier shutdown policy described here considers thresholds $\rho$ as input parameters. In Sections \ref{sec:sol_arch} and \ref{sec:algo} we will describe how to optimize such thresholds. \\

\noindent \textbf{Over-the-top node: Data collection and threshold update.} 
To optimize the load thresholds $\rho=[\rho_{\min},\rho_{\max}]$, defining the carrier shutdown policy implemented at the base station, we use the Over-the-Top (OTT) architecture illustrated in Figure \ref{fig:architecture}. 
At time instants indexed by $t=0,1,\dots$, an OTT computing node retrieves the latest value of the KPI's of interest across the network. 
Then, based on the KPI values, the OTT node is responsible for updating the load thresholds of each sector and pushing the new values to the base stations at appropriate times. Thus, the frequency of threshold update must be lower than or equal to the KPI collection frequency.

As opposed to embedding the solution at the base station, the OTT architecture offers a higher computational power and the ability of having a global view of the network. On the other hand, its bottleneck is represented by the amount of data that can be transferred from the base stations to the OTT node. To cater for this, KPI's are retrieved by the OTT node at (relatively) low frequency, e.g., every 15-60 minutes. This has a decisive impact on the design of our threshold update algorithm, having to deal with a data scarcity issue, as described in the next section.

\section{Load threshold tuning algorithm} \label{sec:algo}

In this section we describe the technical details of the Bayesian algorithm implemented in the OTT node that optimizes the load thresholds $\rho$ for a specific sector.\\

\noindent \textbf{Search region.} 
The load thresholds $\rho=[\rho_{\min},\rho_{\max}]$ can take on any value between $0\%$ and $100\%$, under the condition that $\rho_{\min}<\rho_{\max}$. 
\begin{wrapfigure}{r}{0.27\textwidth}
	\setlength\abovecaptionskip{-0.3\baselineskip}
	\vspace{-1em}
	\begin{center}
		\includegraphics[width=\linewidth]{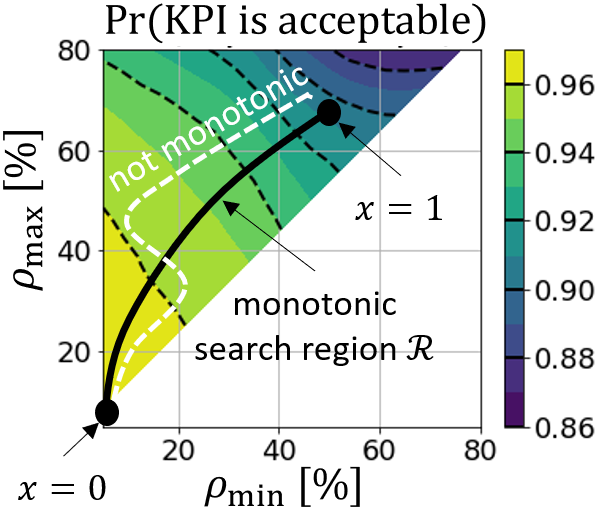}
	\end{center}
	\caption{We consider monotonic threshold search regions $\mathcal R$, along which energy consumption reduces and KPI's degrade.}
	\vspace{-.5em}
	\label{fig:search_region}
\end{wrapfigure}
To simplify our problem, we restrict our threshold search to a restricted region called $\mathcal R$, which we define as a line segment along which both $\rho_{\min}$ and $\rho_{\max}$ are \emph{monotonically non-decreasing}. 
As $\mathcal R$ is one-dimensional, it can be conveniently mapped to a parameter $x\in [0,1]$ such that, as $x$ increases, the corresponding pair $\rho^x=(\rho_{\min},\rho_{\max})$ is element-wise non-decreasing (Fig. \ref{fig:search_region}). 
E.g., $\mathcal R$ can be set to the straight segment between $\rho=[0,0]$ and $\rho=[a,b]$, where $a<b$. In this case, the parameter value $x\in[0;1]$ corresponds to the threshold pair $\rho^x=[xa,xb]$. However, in this paper we do not discuss how one should specifically design $\mathcal R$.\\

\noindent\textbf{Problem reduction.} As the parameter $x$ increases, the expected number of active carriers decreases, since a higher value of $\rho_{\min}$ translates into a higher chance of carrier shutdown, while a higher $\rho_{\max}$ leads to a lower chance of reactivation. 
Hence, as $x$ increases, we can safely assume that the power consumption at the base station reduces \emph{and} that KPI's degrade; in other words, the expectation of the KPI function $f$ decreases. It stems from such considerations that the problem \eqref{eq:es1}-\eqref{eq:es2} under hysteresis policy (Algorithm \ref{alg:LTE1103}) and with load thresholds restricted to $\mathcal R$ boils down to finding the value $x^*$ whose KPI performance is the closest to the target $\xi$:
\begin{align}
	& x^* = \argmin_{x\in [0;1]}   \label{eq:root} \\  
	& \ \ \ \ \Bigg | \lim_{T\rightarrow \infty} \frac{1}{\sum_{t=0}^{T} \!|\mathcal A'_t|} \sum_{t=0}^{T} \sum_{c\in \mathcal A'_t} \!\mathbb E \left[f\left(\{\mathrm{KPI}^{i,c}_t(x)\}_{i=1}^K\right)\right] - \xi \Bigg | \notag
\end{align}
where $\mathrm{KPI}^{i,c}_t(x)$ is the $i$-th KPI value measured at time $t$ in carrier $c$ when the threshold pair $\rho^x$ is under use, and $t$ indexes the instants at which the OTT node collects KPI's. By convention, if there exist multiple solutions to \eqref{eq:root}, then $x^*$ is the highest of them, since it minimizes consumption.

We remark that, if the original problem \eqref{eq:es1}-\eqref{eq:es2} is unfeasible, then \eqref{eq:root} still produces a solution, being the closest one to the feasibility region and such that KPI's are the best possible.\\

\noindent \textbf{Closed-loop paradigm.} To solve \eqref{eq:root} we adopt the following general procedure. At round $k$, upon the selection of value $x_k$ for a specific sector, the carrier shutdown Algorithm \ref{alg:LTE1103} is deployed at the base station with threshold pair $\rho^{x_k}$. 
Then, after a certain time, the resulting KPI values are collected by the OTT node which converts them into binary values---denoted by $\mathcal{D}_k$---via function $f$. Then, a value for $x_{k+1}$ is selected for the next round and the same process is repeated.\\

\noindent \textbf{Vanilla Bayesian algorithm.} We describe our threshold tuning method via a step-by-step approach. We first illustrate the vanilla version of our algorithm under some simplifying assumptions, that we lift in the next paragraphs where the full-blown solution is finally presented. 

We first assume that the binary values $f(\{\mathrm{KPI}^{i,c}_t(x)\}_{i=1}^K)$, measured across different carriers $c\in \mathcal A'_t$ and time instants $t$ and obtained for a specific $x\in[0;1]$, are generated according to an \emph{i.i.d.} Bernoulli random process, where the probability of a sample being 1 is the \emph{unknown} value $p(x)$.
In this case, expression \eqref{eq:root} can be further simplified as follows:
\begin{equation} \label{eq:root_simple}
	x^* = \argmin_{x\in[0;1]} \big|  p(x) - \xi \big|.
\end{equation}
To solve \eqref{eq:root_simple}, one could use the \emph{stochastic approximation} (SA) algorithm that at iteration $k$ chooses a value $x_k$, observes samples $\mathcal{D}_k$ with mean $m_k$, and updates $x$ by a quantity proportional to the excess of $m_k$ with respect to the confidence level $\xi$, i.e., $x_{k+1}=x_k+\epsilon_k (m_k-\xi)$, where $\{\epsilon_k>0\}_k$ must satisfy certain convergence properties \cite{robbins1951stochastic}.

Although it is widely used, its convergence properties are well understood and it requires little computational effort, SA is arguably \emph{not} a good fit for our problem. 
First, it typically converges within few thousands of iterations, which in our case would amounts to a few weeks' time. In fact, one iteration is typically performed every few hours due to the OTT architecture limitations (Section \ref{sec:sol_arch}). 
Moreover, during the first iterations, SA would tend to explore widely across the region $\mathcal R$ before approaching $x^*$, which may cause severe KPI performance drop occurrences. This is clearly unacceptable in most live deployments. Moreover, SA cannot exploit prior information collected via historical data which would help identifying reasonable threshold values from the start. 

For such reasons, we turned our attention towards Bayesian approaches, able to deal with data scarcity and to naturally embed prior information extracted from historical data.\\

\emph{Procedure.} We first parameterize the (unknown) function $p(.)$ as $p_{\theta}(.)$, where $\theta$ are the parameters to be optimized. For instance, $p_{\theta}$ can be defined as a bounded linear function:
\begin{equation} \label{eq:linear}
	p_{\theta}(x) = \min(\max(a - b x, 0),1), \qquad \forall\, x\in[0;1]
\end{equation}
where $\theta=[a,b]$. Our main idea is to compute the most likely values of $\theta$ given the observations and to select the next value of $x$ accordingly. Suppose that at the beginning of iteration $k$ we have a certain probabilistic \emph{belief} on $\theta$, in the form of the probability density $\Pr(\theta)$. Then, the likelihood of observing binary samples $\mathcal{D}_k:=\{d_1,\dots,d_j\}$ given that threshold pair $\rho^{x_k}$ is deployed and that the parameter value is $\theta$, writes:
\begin{align} 
	\Pr(\mathcal{D}_k|\theta) = & \, p_{\theta}(x_k)^{\sum_{i=1}^{J}d_i} \left( 1 - p_{\theta}(x_k)  \right)^{J-\sum_{i=1}^{J}d_i}. \label{eq:es_likelihood} 
\end{align}
The \emph{posterior} belief on $\theta$ is computed via the Bayes rule:
\begin{align}
	\Pr(\theta)\leftarrow \Pr(\theta&|\mathcal{D}_{k}) = \, \frac{\Pr(\mathcal{D}_k|\theta) \Pr(\theta)}{\Pr(\mathcal{D}_k)}, \quad \forall \, k \label{eq:bayes_simple}
\end{align}
where $\Pr(\mathcal{D}_k|\theta)$ is defined in \eqref{eq:es_likelihood}.
In the bounded linear case \eqref{eq:linear} where $\theta$ is two dimensional, \eqref{eq:bayes} can be computed directly via standard numerical techniques. Yet, if $\theta$ has high dimensionality, then computing the denominator of \eqref{eq:bayes} is intractable since it would require the solution of a complex multi-variable integral. In this case, advanced techniques such as Markov Chain Monte-Carlo \cite{brooks2011handbook} are needed. 

Once the belief on $\theta$ is updated, we determine the next value $x_{k+1}$ as the one solving equation \eqref{eq:root_simple} where the true (unknown) value of $p(x)$ is replaced by the expectation of its parametric version $p_{\theta}$ with respect to the updated belief $\Pr(\theta)$, i.e.,
\begin{equation} \label{eq:x_update}
	x_{k+1} = \argmin_{x\in[0;1]} \Big| \mathbb E_{\theta\sim \Pr(\theta)}\left[p_{\theta}(x)\right] - \xi \Big|.
\end{equation}

\begin{figure}[h]
	\setlength\abovecaptionskip{-0.5\baselineskip}
	\vspace{-0em}
	\centering
	\includegraphics[width=.9\linewidth]{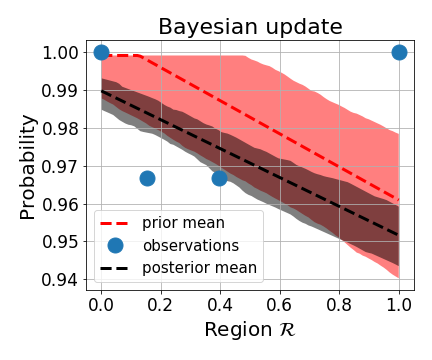}
	\caption{Bayesian update of parameters $\theta$. The bounded linear approximation \eqref{eq:linear} is used. The red and gray shaded regions denote confidence intervals for the value of $p_{\theta}(x)$ with respect to the prior and posterior distribution of $\theta$, respectively. Blue dots are the average of previous observations $\mathcal D$.}
	\vspace{-1.5em}
	\label{fig:es_bayesian}
\end{figure}

\noindent \textbf{Dealing with \emph{long} time-scale traffic variations.} In practice, observed (binary) samples are not \emph{i.i.d.} but they rather follow a distribution that varies along with the traffic characteristics. 
For instance, as the inter-cell interference increases, the KPI's in the sector typically degrade (see Figure \ref{fig:thpt_vs_load_CQI}), which increases the probability of observing a sample equal to 0.

We can account for this in our Bayesian model by assuming that the parameter $\theta:=\theta_k$ varies across iterations $k$ according to a certain Markovian transition law $\Pr(\theta_{k}|\theta_{k-1})$.
In light of this, the Bayes update rule in \eqref{eq:bayes_simple} can be augmented as:
\begin{align}
	\Pr(\theta_k)\leftarrow & \, \Pr(\theta_{k}|\mathcal{D}_{k}) = \frac{\Pr(\mathcal{D}_k|\theta_{k}) \Pr(\theta_k)}{\Pr(\mathcal{D}_k)} \label{eq:bayes} \\
	= & \, \frac{\Pr(\mathcal{D}_k|\theta_{k})\int_{\theta_{k-1}} \Pr(\theta_{k-1}) \Pr(\theta_{k}|\theta_{k-1}) d\theta_{k-1}}{\Pr(\mathcal{D}_k)} \notag 
\end{align}
where the updated belief $\Pr(\theta_{k})$ is written as the convolution between the former belief $\Pr(\theta_{k-1})$ and the transition rule $\Pr(\theta_{k}|\theta_{k-1})$. If the parameter $\theta$ is static, then $\Pr(\theta_k|\theta_{k-1})=\ind(\theta_k=\theta_{k-1})$ and we recover the original update \eqref{eq:bayes_simple}.

We remark that the transition rule $\Pr(\theta_k|\theta_{k-1})$ is unknown but there exist techniques (e.g., \cite{west2013bayesian}) to estimate it from data. \\


\noindent \textbf{Dealing with \emph{short} time-scale traffic variations.} The technique described is able to effectively track the changes in the distribution of $\theta$ when they occur on a relatively slow time scale, in the order of a few iterations.
\begin{wrapfigure}{r}{0.3\textwidth}
	\setlength\abovecaptionskip{-0.3\baselineskip}
	\vspace{-1em}
	\begin{center}
		\includegraphics[width=\linewidth]{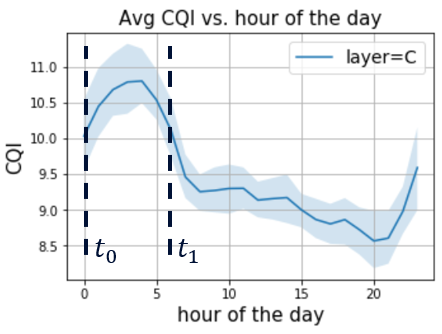}
	\end{center}
	\caption{Two ($N=2$) windows are identified here via \eqref{eq:psw_opt}. Shaded blue region is the confidence interval for CQI distribution.}
	\vspace{-.5em}
	\label{fig:psw_opt}
\end{wrapfigure}
However, a single iteration may span several hours, during which traffic may follow typical peaks and troughs causing abrupt temporal changes to the distribution of $\theta$ over temporal scales not accounted for in the above approach.
To tackle this, a practical shortcut is to pre-emptively split the 24 hours of the day into $N$ windows during which traffic conditions are typically stable, and run independent Bayesian update instances on each window. 
For a given window, thresholds can be updated on a daily basis. Therefore, window splitting caters for short-time scale traffic variations within a single day, 
while the transition law $\Pr(\theta_k|\theta_{k-1})$ deals with long-term variations,
across multiple days. Such $N$ windows can be then defined, e.g., as those during which CQI is the most stable, i.e.,
\begin{equation} \label{eq:psw_opt}
	\min_{N,t_0<t_1<\dots<t_{N-1}} \frac{1}{N}\sum_{i=0}^{N-1} \mathrm{Std}(\mathrm{CQI}[h_i,h_{\mathrm{mod}(i+1,N)}])
\end{equation}
where $\mathrm{Std}(\mathrm{CQI}[h_i,h_{i+1}])$ is the empirical standard deviation of CQI values within the hours of the day $[h_i,h_{i+1}]$, computed on historical data collected in the sector to be optimized. 

As windows get shorter, the amount of KPI data collected at each iteration reduces, which bears a negative impact on the convergence properties of our Bayesian approach. Thus, it is important to ensure a minimum duration of a few hours for each window, that can be added as a constraint to \eqref{eq:psw_opt}.

\begin{algorithm}
	\KwIn{Search region $\mathcal R$} 
	Split the 24 hours into $N$ windows via \eqref{eq:psw_opt}\\
	\For{\textnormal{window} $n=1,\dots,N$}
	{
	Collect historical data and initialize prior $\Pr(\theta_0)$\\
	\For{\textnormal{day} $k=1,2,\dots$}{
		Compute $x_{k}$ via \eqref{eq:x_update}\\
		Deploy load thresholds $\rho^{x_{k}}$ and collect KPI's\\
		Compute $\Pr(\theta_k)$ via \eqref{eq:bayes} 
	}}
	\caption{Load threshold tuning algorithm}
	\label{alg:thrs_tuning}
\end{algorithm}


\noindent \textbf{Prior initialization.} In order to accelerate the convergence speed of the Bayesian search and avoid a cold start, it is good practice to properly initialize the \emph{prior} belief $\Pr(\theta_0)$, \emph{before} the online exploration phase begins \cite{maggi2021bayesian}. 
First, by construction of the search region $\mathcal R$, we know that $p_{\theta}(x)$ is a \emph{non-increasing} function of $x$. Thus, we start by assigning a null probability to all values $\theta$ for which the monotonicity condition is not verified. 
The prior belief on $\theta$ can be also refined via historical data---obtained from live network deployments or from simulation---reporting the KPI's of interest obtained for different values of load thresholds within $\mathcal R$. Then, the Bayes update \eqref{eq:bayes} is performed for each of the historical threshold values, as if the algorithm ``discovered'' them in online fashion.

Our threshold tuning procedure is resumed in Algorithm \ref{alg:thrs_tuning}.

\section{Live network trials}

We tested our solution for carrier shutdown in a proof of concept (PoC) on a live customer 4G network, over a cluster comprising 19 sites (and 57 sectors). Most of the sites had 4 frequency layers (800, 1800, 2100 and 2600 MHz). 
\emph{Baseline} measurements were taken over periods spanning a few weeks
immediately before and after the PoC trial, during which all carriers were kept active. Note that this corresponds to the extreme case where $\rho=[0,0]$. Two ($N=2$) windows were identified for each sector, one during daytime and the other during nighttime. 
The bounded linear parameterized function \eqref{eq:linear} was used. The prior $\Pr(\theta_0)$ was initialized by collecting 2 weeks data during the baseline period. Multiple instances of the threshold tuning algorithm were running in an OTT server for a duration of 4 weeks, where each instance optimized thresholds for a specific sector and window. 
The search region included the origin $\rho=[0,0]$, hence guaranteeing the possibility to replicate the baseline behavior if needed. The parameter transition rule $\Pr(\theta_k|\theta_{k-1})$ was set to a Gaussian distribution with zero mean and diagonal covariance matrix, which allowed the algorithm to adapt to traffic variations by gradually ``forgetting'' past observations. 
Remarkably, each Bayesian update \eqref{eq:bayes} could be computed in less than 1 second. We chose the IP downlink throughput in QCI 8 as the KPI to be preserved, with an associated target of $y=5$ Mbps and confidence level $\xi=89\%$. To preserve coverage, 800 and 1800 MHz carriers were always left active. 

\begin{figure}[h]
	
\centering
\setlength\abovecaptionskip{0.2\baselineskip}
		\subfloat{
			\includegraphics[width=0.5\textwidth]{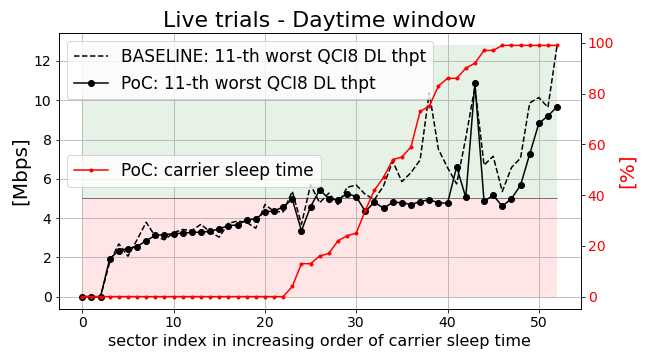}
			\label{sub:fig1}
		} \\
		\subfloat{
			\includegraphics[width=0.5\textwidth]{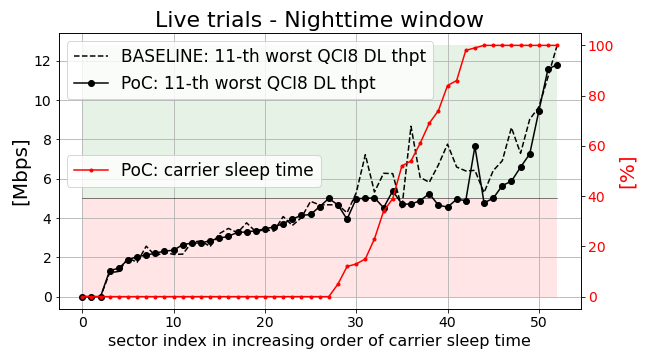}
			\label{sub:fig2}
		}
		\setlength{\belowcaptionskip}{-2cm}
		\caption{Live network PoC results. We show the 11-th worst QCI8 DL throughput value and the carrier sleep time (i.e., the percentage of time that carriers eligible for shutdown are actually deactivated) for each sector. We benchmark against the ``baseline'' situation where all carriers are active.}
		\vspace{-1.5em}
		\label{fig:trial_results}

\end{figure}

During our PoC, we could reduce the energy consumption at the base station by 11$\%$ with respect to baseline, which is a significant given that energy is up to $40\%$ of an operator's OPEX \cite{gsma}. Overall, carriers were shut down for around 30$\%$ of the time. We detected no significant impact on cell congestion, PDCP traffic volume, or number of active users, neither on the cluster of optimized sites nor on neighboring ones. 
Figure \ref{fig:trial_results} shows that our main principle \eqref{eq:root} for energy savings was satisfied. Indeed, in the sectors where KPI was violating the constraint (i.e., the 11-th worst KPI value was lower than 5 Mbps) even in the baseline phase, no carriers were (rightly) ever shut down during the PoC. Conversely, for the sites where KPI's was above the target, carriers were put to sleep at a rate guaranteeing the KPI to meet the constraint with approximate equality. For a few sectors, KPI's were still above target even if all carriers---among those eligible for shutdown---were sleeping all the time.

\section{Conclusions}

By shutting carriers down, the power consumption at the base station can be significantly reduced. However, this comes with the cost of degrading the user quality of service. 
We designed a practical solution that minimizes the power consumption at the base station while guaranteeing that pre-selected KPI's are acceptable with high confidence. A carrier shutdown policy depending on some threshold parameters is implemented at the base station. An over-the-top node optimizes the thresholds via a data efficient Bayesian procedure. 
During live networks trials our method could reduce the power consumed by the base stations by 11$\%$ while fulfilling the KPI constraints in each sector.

\bibliographystyle{IEEEtran}
\bibliography{Literature_lorenzo}

\end{document}